\documentclass[11pt,fleqn,twoside]{article}
\usepackage{amsfonts,amssymb,latexsym}
\makeatletter
\newcommand{\prava}[1]{\small\it
\begin{flushleft}
Copyright \copyright \ 1999 by  #1
\end{flushleft}}

\newcommand{\name}[1]{\begin{flushleft}
                       \LARGE \bf #1
                       \end{flushleft}\vspace{-3mm}}

\newcommand{\Author}[1]{\begin{flushleft}
                       \it #1 \end{flushleft}}

\newcommand{\Adress}[1]{\begin{flushleft}
                       \it #1 \end{flushleft}}

\newcommand{\Date}[1]{\begin{flushleft}
                      \small  \it #1 \end{flushleft}}

\newcommand{\ehkol}{Author \ name}
\newcommand{\ohkol}{Article \ name}
\renewcommand{\@evenhead}{
\hspace*{-3pt}\raisebox{-15pt}[\headheight][0pt]{\vbox{\hbox to \textwidth 
{\thepage \hfil \ehkol}\vskip4pt \hrule}}}
\renewcommand{\@oddhead}{
\hspace*{-3pt}\raisebox{-15pt}[\headheight][0pt]{\vbox{\hbox to \textwidth 
{\ohkol \hfil \thepage}\vskip4pt\hrule}}}
\renewcommand{\@evenfoot}{}
\renewcommand{\@oddfoot}{}

\newcommand{\be}{\begin{equation}}
\newcommand{\ee}{\end{equation}}
\newcommand{\ba}{\hspace*{-5pt}\begin{array}}
\newcommand{\ea}{\end{array}}

\newcommand{\ds}{\displaystyle}
\makeatother

\def\ve#1{\mid #1\rangle}
\def\vc#1{\langle #1\mid}

\renewcommand{\theequation}{\thesection.\arabic{equation}}

\begin{document}

\thispagestyle{empty}
\setcounter{page}{181}

\renewcommand{\ehkol}{A.N. Leznov}
\renewcommand{\ohkol}{Exactly Integrable Systems Connected to Semisimple Algebras}

\begin{flushleft}
\footnotesize \sf
Journal of Nonlinear Mathematical Physics \qquad 1999, V.6, N~2,
\pageref{leznov-fp}--\pageref{leznov-lp}.
\hfill {\sc Article}
\end{flushleft}

\vspace{-5mm}

{\renewcommand{\footnoterule}{}
{\renewcommand{\thefootnote}{} \footnote{\prava{A.N. Leznov}}}

\name{Exactly Integrable Systems Connected\\
to Semisimple Algebras of Second Rank \\
{\mathversion{bold} $A_2$, $B_2$, $C_2$, $G_2$}} \label{leznov-fp}

\Author{A.N. LEZNOV~$^{a,b,c}$}

\Adress{$^{(a)}$~IIMAS-UNAM, Apartado Postal 20-726, Mexico DF 01000,
Mexico\\[1mm]
$^{(b)}$~Institute for High Energy Physics, 142284
Protvino, Moscow Region, Russia\\[1mm]
$^{(c)}$~Bogoliubov Laboratory of Theoretical Physics, JINR,\\
$\phantom{^{(c)}}$~141980 Dubna, Moscow Region, Russia\\[1mm]
E-mail: leznov@ce.ifisicam.unam.mx }

\Date{Received October 10, 1998; Revised November 27, 1998; Accepted November 30, 1998}

\begin{abstract}
\noindent
 Exactly integrable systems connected  to semisimple algebras
of second rank with an ar\-bitrary choice of grading are
presented in explicit form. General solutions of these systems are expressed in
terms of matrix elements of two fundamental representations of
the corresponding semisimple groups.
\end{abstract}

\renewcommand{\thefootnote}{\arabic{footnote}}
\setcounter{footnote}{0}

\section{Introduction}

The main goal of this paper is to demonstrate, on the examples of
semisimple algebras of second order ($ A_2, B_2, C_2, G_2$),
 the general
construction connecting a semisimple algebra of a given grading to an
exactly integrable system.
The simplest example is the two-di\-men\-sional Toda lattice considered and
integrated in the case of an arbitrary semisimple algebra almost 20 years ago
\cite{leznov:LG, leznov:ls0}\footnote{For  $A_n$ series this problem was solved more then 150 years
ago in Darbouxs  papers!}. For
main grading, exactly integrable systems were explicitly found and described in
the recent papers of the author \cite{leznov:LM} (so called Abelian~case).

In the present paper we follow three dif\/ferent and independent aims.
The f\/irst is to relate unknown up to now integrable systems to
nonabelian gradings\footnote{Note that the zero order subspace is a non-commutative
algebra by itself.} (see \cite{leznov:GC} in this con\-nec\-tion).
The second one is to get rid of the restriction of nonabelian Toda theory
to use only subspaces with zero and $\pm 1$ grading indices. The last, but not
the least important one, is to provide the reader with a scheme of how the group
representation theory (in the very restricted volume) can be applied   to the
theory of integrable systems.

For our purposes here it is not the shortest and simplest way to the result that
is important, but the result by itself. Therefore, in concrete examples we
tried to use calculations that can be followed and checked directly using only
simplest algebra.

The paper is organized as follows. Section~2 contains the background information
on representation theory of semisimple algebras and groups (as a rule without
proofs). Section~3 describes the general construction, mathematical tricks and
methods used in main sections. In Section~4 concrete examples of semisimple
algebras of second order are considered in details for all possible
gradings. Concluding remarks and perspectives for further investigation
are outlined in Section~5.

\section{Semisimple algebras and groups}

Let ${\cal G}$ be an arbitrary f\/inite-dimensional graded Lie
algebra\footnote{We make no distinction between algebras and super-algebras,
just keeping in mind that even (odd) elements of super-algebras are always
multiplied by even (odd) elements of the Grassman space.}.
Then $\cal G$ can be written as a direct sum of subspaces of dif\/ferent grading
indices
\begin{equation}
  {\cal G}=\left(\oplus^{N_-}_{k=1} {\cal G}_{-\frac{k}{2}}\right)
  {\cal G}_0 \left(\oplus^{N_+}_{k=1}{\cal G}_{\frac{k}{2}}\right).
\label{GR}
\end{equation}

Generators with an integer grading index are called bosonic, while those with
half-integer grading index are named fermionic. The positive (negative) grading
corresponds to upper (lower) triangular matrices.

The grading operator $H$ for an arbitrary semisimple algebra can be written
as a linear combination of elements of commutative Cartan subalgebra taking
unity or zero values on the generators of simple roots 
\begin{equation}
H=\sum^r_{i=1} \left(K^{-1}c\right)_i h_{i}.
\label{cartan1}
\end{equation}
Here $K^{-1}$ is the inverse Cartan matrix $K^{-1}K=KK^{-1}=I$
and $c$ is a column of zeros and unites in an arbitrary order.
Under the main grading all $c_i=1$. In this case
$(K^{-1}c)_i=\sum\limits_{j=1}^r K^{-1}_{i,j}$, where $r$ is the rank of the algebra.

As usually, generators of simple roots $X^{\pm}_i$ (raising/lowering
operators) and Cartan elements $h_i$  satisfy the system of commutation
relations:
\begin{equation}
[h_i , h_j]=0, \qquad [h_i,X^{\pm}_j]=\pm K_{j,i}X^{\pm}_j, \qquad
[X^{+}_i,X^{-}_j\}={\delta}_{i,j} h_j, \quad (1 \leq i,j \leq r),\label{aa6}
\end{equation}
where $K_{ij}$ are elements of Cartan matrix and brackets $[,\}$
stand for the graded commutator.

The highest vector $\ve{j}$ ($\vc{j} \equiv \ve{j}^{\dagger}$) of
the $j$--th fundamental representation has the following properties:
\be
X^{+}_i\ve{j}=0, \qquad h_i\ve{j}={\delta}_{i,j}\ve{j}, \qquad \vc{j}\ve{j}=1.
\label{high}
\ee
The representation is exhibited by applying
lowering operators $X^{-}_i$ to the vector $\ve{j}$ repeatedly and
extracting all linearly-independent vectors with non-zero norm. The f\/irst
few basis vectors are
\be
\ve{j}, \quad X^{-}_j\ve{j},  \quad X^{-}_i X^{-}_j\ve{j}\neq 0,\quad
K_{i,j}\neq 0,\quad i\neq j.
\label{vectors}
\ee
}

In fundamental representations an important identity for matrix elements of a
group element $G$ holds\footnote{Recall that a
superdeterminant is def\/ined as $\mbox{sdet} \left(\begin{array}{cc} A, & B \\ C, & D
\end{array}\right) \equiv \det (A-BD^{-1}C ) (\det D)^{-1}$.}~\cite{leznov:ls0}
\be
\mbox{sdet} \left(\begin{array}{cc} \vc{j}X_j^+GX_j^-\ve{j}, &
\vc{j}X_j^+G\ve{j}
\vspace{3mm}\\
\vc{j}GX_j^-\ve{j}, & \vc{j} G \ve{j} \end{array}\right) = \prod^r_{i=1}\vc{i} G \ve{i}^{-K_{ji}},
\label{recrel}
\ee
 The identity (\ref{recrel}) is in fact a generalization (to the case of an arbitrary semisimple
Lee super-group) of the famous Jacobi identity that relates
determinants of orders $(n-1)$, $n$ and $(n+1)$  of some special
matrices. As we will see in the next section, this identity is of
such importance in deriving exactly integrable systems that one
can even say that it is responsible for their existence. We will
still refer to  (\ref{recrel}) as to ``the f\/irst Jacobi
identity''. In addition to (\ref{recrel}), there exists another
independent identity of key importance~\cite{leznov:l}
\be
\ba{l} \ds (-1)^P K_{i,j} {\vc{j}X_j^+X_i^+ G \ve{j}\over \vc{j} G
\ve{j}}+ K_{j,i} {\vc{i}X_i^+X_j^+ G \ve{i}\over \vc{i} G \ve{i}}
\vspace{3mm}\\ \ds \qquad +K_{ij}K_{j,i}(-1)^{jP}{\vc{j}X_j^+ G
\ve{j}\over \vc{j} G \ve{j}}{\vc{i}X_i^+ G \ve{i}\over \vc{i} G
\ve{i}}=0 ,\qquad i\neq j \ea\label{J2} \ee which will be called
the second Jacobi identity. This identity is responsible (in the
above sense) for the existence of hierarchies of integrable
systems invariant with respect to integrable mappings that are
connected to every exactly integrable system.

Either from (\ref{recrel}) or from (\ref{J2}) it is possible to construct
many usefull recurrent relations that are used in further
consideration.

Taking into account the importance of Jacobi identities (\ref{recrel})
and (\ref{J2}) for further consideration we
present below a brief proof of (\ref{recrel}).

Let us consider the left hand side of (\ref{recrel}) as a function on the
group. The action on an arbitrary group element $G$ in the def\/inite
representation $l$ of the operators of the right (left) regular
representation is by def\/inition
\begin{equation}
M_{\mbox{\scriptsize left}}(\tilde M_{\mbox{\scriptsize right}}) G= M_l G (\tilde M_l),
\label{AM}
\end{equation}
where $M_l$, $\tilde M_l$ are the generators (the matrices of
corresponding dimension) of shifts on the group in a given $l$
representation. Now let us act with an arbitrary generator of the
simple positive root $(X^+_s)_r$ on the left hand side of
(\ref{recrel}). This action is equivalent to dif\/ferentiation and
therefore should be applied consequently to the f\/irst and second
columns of the matrix (\ref{recrel}) adding the results. The
action on the second column results in zero as a corollary of the
def\/inition of the higest state vector (\ref{high}). Action on
the f\/irst column is dif\/ferent from zero only in the case
$s=j$. But in this case using the same def\/inition of the highest
state vector we conclude that as a result of dif\/ferentiation of
the f\/irst column it becomes equal to the second one with the
zero f\/inal result. Thus considered as a function on the group
the left hand side of (\ref{recrel}) is also proportional to the
highest vector (or a linear combination of such vectors) of some
other representation. The higest vector of the irreducible
representation is uniquely def\/ined by the values that Cartan
generators take on it. If Cartan generators take on the highest
vector values $V(h_i)=l_i$, the last can be uniquely represented
in the form
\begin{equation}
\vc{l} G \ve{l}=C \prod_{i=1}^r(\vc{i} G \ve{i})^{l_i}.\label{HVD}
\end{equation}
Calculating the values of Cartan generators on
the left hand side of equation (\ref{recrel}) (both left and right with
the same result) and using the last comment
about the form of the highest vector, we prove (\ref{recrel}) ($C=1$, as can be
seen by putting $G=1$ and comparing both sides).

The second Jacobi identity can be proven by similar argument \cite{leznov:l}.

The following generalization of the f\/irst Jacobi identity will be very
important in calculations dealing with nonabelian gradings.

Let $\ve{\alpha}$ be basis vectors of some representation in the
strict order of increasing the number of lowering generators (see
(\ref{high}) and (\ref{vectors})). We also assume that the action
of a generator of an arbitrary positive simple root on each basis
vector results in a linear combination of the previous ones.

Then the principal minors of an arbitrary order of the matrix
($G$ is an arbitrary element of the group):
\[
G_{\alpha}=\vc{\alpha} G  \ve{\alpha}
\]
are annihilated from the right (from the left) by generators of positive
(negative) roots.

Indeed this is equivalent to dif\/ferentiation and therefore it is
necessary to act on each column (line) of the minors matrix and
add the results. But the action of the generator of a positive
simple root on the state vector with a given number of lowering
operators transforms it into a state vector with a number of
lowering operators on unity less, which according to our
assumption is a linear combination of previous columns (lines).
Thus in all cases the lines or columns of the resulting
determinant are linearly dependent with zero result.

The generators of Cartan subalgebra obviously take the def\/inite values on
minors of these kind and if the corresponding values are
$l^s_i$, it is possible to write the equality in correspondence with (\ref{HVD})
\begin{equation}
Min_s= C_s \prod_{i=1}^r \vc{i} G \ve{i}^{l^s_i},\label{GJI}
\end{equation}
where constants $C_s$ can be determined as described above.

\setcounter{equation}{0}

\section{General construction and technique of computation}

The grading of a semisimple algebra is def\/ined by the values that the grading
operator $H$ takes on the simple roots of the algebra. As it was mentioned
above, this values can be only zeros and unites in an arbitrary order.
\[
[H, X^{\pm}_i]=\pm X^{\pm}_i,\qquad H=\sum_1^r (K^{-1}c)_i h_i,\qquad c_i=1,0.
\]

On the level of Dynkin's diagrams the grading can be introduced by
using two colors for its dots: black for simple roots with $c_i=1$
and red for roots with $c_i=0$. To each consequent sequence of the
red (simple) roots the corresponding semisimple algebra
(subalgebra of the initial one) is connected. All these algebras
are obviously mutually commutative and belong to the zero graded
subspace.  Cartan elements of the black roots also belong to the
zero graded subspace. We will use the usual numeration of the dots
of Dynkin diagrams and all red algebras will be distinguished by
an index of their f\/irst root~$m_s$. The rank of $m_s$-th red
algebra will be denoted as $R_s$. Thus
$X^{\pm}_{m_s},X^{\pm}_{m_s+1},\ldots,X^{\pm}_{m_s+R_s-1}$ is the
system of simple roots of $m_s$ red algebra.

After these preliminary comments turn  to the general
construction \cite{leznov:l1}.

Let two group valued functions $ M^+(y)$, $M^-(x)$ be solutions of $S$-matrix
type equations
\begin{equation}
\ba{l}
\ds M^+_y=\left(\sum_0^{m_2} B^{(+s}(y)\right) M^+\equiv\left(B^{(0}+L^+\right)M^+,
\vspace{3mm}\\
\ds M^-_x=M^-\left(\sum_0^{m_1} A^{(-s}(x)\right)\equiv M^-\left(A^{(0}+L^-\right),
\ea
 \label{I}
\end{equation}
where $B^{(+s}(y)$, $A^{(-s}(x)$ take values  in $\pm s$ graded
subspaces correspondingly and $s=0, 1, 2,\ldots,m_{1,2}$. In each f\/inite-dimensional
representation  $B^{(+s}(y)$, $A^{(-s}(x)$ are upper (lower) triangular matrices and therefore
equations~(\ref{I}) are integrated in quadratures.

The composite group valued function $K$ plays the key role in our construction
\begin{equation}
K=M^+ M^-. \label{II}
\end{equation}
It turns out that matrix elements of $K$ in various fundamental
representations are related by closed systems of equivalent relations,
which can be interpreted as exactly integrable system with known general
solution.

Bellow we describe calculation methods to prove
this proposition.

First of all let us calculate the second mixed derivative $(\ln \vc{i} K
\ve{i})_{x,y}$, where index $i$ belongs to the black dot of Dynkin diagram.
We have
\begin{equation}
(\ln \vc{i} K \ve{i})_x= {\vc{i} K (A^0+L^-) \ve{i}\over \vc{i} K \ve{i}}=
A^0_i(x)+ {\vc{i} K L^- \ve{i}\over \vc{i} K \ve{i}}.
\label{MI}
\end{equation}
Indeed, $ K_x=M^+(y)M^-_x(x)=K (A^0+L^-)$ as a corollary of equation for
$M^-$. All red components of $A^0$ under the action on the black highest
vector state $ \ve{i}$ lead to zero result in connection with
(\ref{vectors}).
The action of Cartan elements of the black roots state vector satisf\/ies
the condition $h_j \ve{i}=\delta_{i,j} \ve{i}$ and thus only coef\/f\/icient on
$h_i$ remains in the f\/inal result (\ref{MI}).

Further dif\/ferentiation (\ref{MI}) with respect to $y$, with the
help of arguments above, leads to following result:
\begin{equation}
(\ln \vc{i} K \ve{i})_{x,y}=\vc{i} K \ve{i})^{-2} \pmatrix{
\vc{i} K \ve{i}, & \vc{i} K L^- \ve{i} \cr
        \vc{i} L^+ K \ve{i}, &  \vc{i} L^+ K L^- \ve{i} \cr}.
\label{AR}
\end{equation}

Applying (\ref{AM}) of the previous section to the left hand side of (\ref{AR}), we f\/inally obtain
\begin{equation}
(\ln \vc{i} K \ve{i})_{x,y}=L^-_rL^+_l \ln \vc{i} K \ve{i})^{-1}.
\label{ARR}
\end{equation}

Thus the problem of calculating  the mixed second derivative is
reduced to purely algebraic manipulations on the level of representation theory
of semisimple algebras and groups. Further evaluation of (\ref{ARR}) is
connected with repeated application of the f\/irst~(\ref{recrel})
and second (\ref{J2}) Jacobi identities as it will be clear from the material
of the next section.

As it was mentioned above, the red algebras of zero order graded subspace in
general case are not commutative. This leads to additional
computational dif\/f\/iculties. Let us denote by $ \ve{m_i}$
the highest vector of $m_i$th fundamental representation of the initial
algebra. Of course, $\ve{m_i}$ is simultaneously the highest vector of the
f\/irst fundamental representation of the $m_i$ red algebra. Let $\vc{\alpha_i},
\ve{\beta_i}$ be basis vectors of the f\/irst fundamental representation
(this restriction is not essential) of $m_i$-th red algebra and let us
consider the matrix elements of element $K$ in this basis.
$R_i+1\times R_i+1$ matrix ($R_i+1$ is the dimension of the f\/irst fundamental
representation) with matrix elements  $\vc{\alpha_i} K \ve{\beta_i}$
will be denoted by a single symbol $u_i$ (index $i$ takes values from one to
the number of the red algebras, which is the function of the choosen grading).

For derivatives of matrix elements of so constructed matrix we have
consequently (index~$i$ we omite for a moment):
\begin{equation}
\vc{\alpha} u_x \ve{\beta}=\vc{\alpha} K (A^0+L^-) \ve{\beta}=\! \sum_
{\gamma} \vc{\alpha} K  \ve{\gamma}\vc{\gamma} I A^0 \ve{\beta}
+\vc{\alpha} K L^- \ve{\beta}. \label{MCD}
\end{equation}
Or equivalently
\[
u^{-1} u_x= A^0(x)+u^{-1} \vc{} K L^- \ve{}.
\]
Further dif\/ferentiation with respect to $y$ variable leads to
\be
\ba{l}
\ds \vc{}((u^{-1} u_x)_y\ve{}=u^{-1} \vc{} (B^0+L^+) K L^- \ve{}-u^{-1}\vc{}
(B^0+L^+) K \ve{} u^{-1}\vc{} K L^-\ve{}
\vspace{2mm}\\
\ds \qquad = u^{-1}(\vc{} L^+ K L^- \ve{}-\vc{} L^+ K \ve{} u^{-1} \vc{} K L^-\ve{}).
\ea \label{MC}
\end{equation}
The last expression  may be brought to the form of the ratio of
two determinants  of $R_i+2$ and $R_i+1$ orders respectively with
the help of standard transformations:
\begin{equation}
\vc{} u (u^{-1} u_x)_y \ve{}={\mbox{Det}_{N_i+1}\pmatrix{ u & K L^-\ve{} \cr
\vc{} L^+ K  & \vc{} L^+ K  L^- \ve{} \cr}\over \mbox{Det}_{N_i}(u)}. \label{MCC}
\end{equation}

The generalised Jacobi identity (\ref{GJI}) of the previous section plays the key role for
discovery of the last expression and will be exploited many times.

\setcounter{equation}{0}

\section{The algebras of second rank {\mathversion{bold}$A_2$, $B_2, C_2$, $G_2$}}

All elements of these algebras may be constructed by consequent
multi-commutation of generators of four simple roots $X^{\pm}_{1,2}$ with the
basic system of commutation relations
\be\label{RS}
\ba{l}
[X^+_1,X^-_1]=h_1,\qquad [X^+_1,X^-_2]=[X^+_2,X^-_1]=0,\qquad[X^+_2,X^-_2]=h_2,
\vspace{2mm}\\
\ds {}[h_1,X^{\pm}_1]= {\pm} 2X^{\pm}_1, \qquad [h_2,X^{\pm}_2]= {\pm} 2X^{\pm}_2,
\vspace{2mm}\\
\ds {}[h_1,X^{\pm}_2]= {\mp} pX^{\pm}_2, \qquad [h_2,X^{\pm}_1]= {\mp} X^{\pm}_1,
\qquad p=1,2,3.
\ea
\ee
In all cases there are three possible nontrivial gradings: $(1,1)$ -- the
principle one (Abelian case), $(1,0)$ -- the grading of the f\/irst simple root
and $(0,1)$ -- of the second simple one. In the case of the principle grading
corresponding integrable systems for arbitrary semisimple algebras were found
and described in \cite{leznov:LM}. Each further subsections will be
devoted to detail consideration of nonabelian gradings $(1,0)$, $(0,1)$, which
are equivalent to each other only in the case of $A_2$ algebra.

In the end of this mini-introduction we present the second Jacobi identity
as applied to the algebras of second rank:
\begin{equation}
{\vc{2} X^+_2 X^+_1 K \ve{2}\over \vc{2} K \ve{2}}+p {\vc{1} X^+_1 X^+_2 K
\ve{1}\over \vc{1} K \ve{1}}=p {\vc{2} X^+_2  K \ve{2}\over \vc{2} K \ve{2}}
{\vc{1} X^+_1 K \ve{1}\over \vc{1} K \ve{1}}\label{2JI}
\end{equation}
or in notation, which will be introduced by the way of consideration
\[
\bar \alpha_{21}+p\bar \alpha _{12}=p\bar \alpha _1\bar \alpha _2,\qquad
\alpha_{12}+p\alpha _{21}=p\alpha _1 \alpha _2.
\]

\subsection{Unitary {\mathversion{bold}$A_2$} serie}

The root system of this algebra consists of three elements with the
generators $X^{\pm}_1,X^{\pm}_1,X^{\pm}_{12}$ $\equiv \pm [X^{\pm}_1,X^{\pm}_2]$.
This case corresponds to $p=1$ in (\ref{RS}). For def\/initeness we restrict
ourselves by $(1,0)$ grading $[H,X^{\pm}_1]=\mp X^{\pm}_1$, $[H,X^{\pm}_2]=0$.

$L^{\pm}$ operators belong to $\pm 1$ graded subspaces and have the form:
\[
L^+=\bar c_1 X^+_1+\bar c_2 [X^+_2,X^+_1],\qquad L^-=c_1 X^-_1+c_2 [X^-_1,X^-_2],
\]
where $c_{1,2}\equiv c_{1,2}(x)$, $\bar c_{1,2}\equiv \bar c_{1,2}(y)$.

The object of investigation is $2\times 2$ matrix $u$ in the basis of the
second fundamental representation of $A_2$ algebra\footnote{The (bra) basis
vectors of the three dimensional (``qwark'') second fundamental representation
of $A_2$ algebra are the $\vc{2}$, $\vc{2} X^+_2$, $\vc{2} X^+_2 X^+_1 $.}:
\be
u=\left(\begin{array}{cc} \vc{2} K \ve{2}, & \vc{2} K X_2^-\ve{2} \vspace{1mm}\\
\vc{2}X_2^+ K \ve{2}, & \vc{2}X_2^+ K X_2^-\ve{2}
\end{array}\right). \label{recrelI}
\ee
In correspondence with (\ref{MCC}) we have:
\begin{equation}
\vc{} u (u^{-1} u_x)_y \ve{}={\mbox{Det}_3\pmatrix{ u & I K L^- \ve{} \cr
                   \vc{} L^+ K I & \vc{} L^+ K  L^- \ve{} \cr}\over \mbox{Det}_2(u)}.
\label{MCC"}
\end{equation}
The action of operators $L^{\pm}$ on basis vectors $\ve{2}, X^-_2 \ve{2}$
($\vc{2},\vc{2} X^+_2$) is the following:
\[
\ba{l}
L^-\ve{2}=c_2 X^-_1 X^-_2 \ve{2}, \qquad L^-X^-_2\ve{2}=c_1 X^-_1 X^-_2 \ve{2},
\vspace{2mm}\\
\vc{2}L^+=\bar c_2 \vc{2} X^+_2 X^+_1,\qquad  \vc{2} X^+_2 L^+=\bar c_1
\vc{2} X^+_2 X^+_1.
\ea
\]
So in this case the following sequence of basis vectors from generalized
Jacobi identity (\ref{GJI}) takes places:
\[
\vc{2}, \quad \vc{2} X^+_2, \quad \vc{2} X^+_2 X^+_1.
\]
The summed values of Cartan generators $h_1$, $h_2$ on this basis take
zero values and so $\mbox{Det}_3$ from (\ref{MCC"}) equal to unity (with correct
account of the constant). This is a really highest vector of scalar,
one-dimensional representation of $A_2$ algebra.

Finally (\ref{MCC"}) leads to the system, which matrix function $u$ satisfy:
\begin{equation}
(u^{-1} u_x)_y =(\mbox{Det}\; u)^{-1} u^{-1}\pmatrix{ c_2 \bar c_2, & c_1 \bar c_2  \cr
                   c_2 \bar c_1, & c_2 \bar c_2 \cr}.
\label{A_2}
\end{equation}

In usual notations the system (\ref{A_2}) is nonabelian $A_2 (1,0)$ Toda
chain. The system (\ref{A_2}) is obviously form-invariant with respect to
transformation:
\[
u\to \bar g(y) u \bar g(x).
\]
With the help of this transformation the arbitrary up to now functions
$c$, $\bar c$ may be evaluated to a constant values.

\subsection{Orthogonal {\mathversion{bold}$B_2$} serie equivalent to simplectic one
{\mathversion{bold}$C_2$}}

This case corresponds to the choise $p=2$ in (\ref{RS}). Both
gradings are not equivalent to each other and must be considered
separately. First fundamental representation for $B_2$ algebra is
the second one for $C_2$ serie and vice versa.

\subsubsection{(1,0) grading}

Generators $L^{\pm}$ may contain components with $\pm 1$, $\pm2$ graded
indexes and have the form:
\[
\ba{l}
L^+=\bar c_1 X^+_1+\bar c_2 [X^+_2,X^+_1]+\bar c^2 [[X^+_2,X^+_1]X^+_1],
\vspace{2mm}\\
L^-=c_1 X^-_1+c_2[X^-_1,X^-_2]+c^2 [ X^-_1[X^-_1,X^-_2]].
\ea
\]
The object of investigation is two dimensional matrix $u$ in the basis of
the second fundamental representation of $B_2$ algebra. The main equation
(\ref{MCC"}) also does not change. The action of $L^{\pm}$ operators on
the basis vectors have now the form\footnote{Five basis vectors of the f\/irst fundamental
representation of the $B_2$ algebra are the following: $ \ve{2}$, $X^-_2 \ve{2}$,
$X^-_1 X^-_2 \ve{2}$, $X^-_1 X^-_1 X^-_2 \ve{2}$, $X^-_2 X^-_1 X^-_1 X^-_2 \ve{2}$.}:
\[
\ba{l}
L^-\ve{2}=(c_2+c^2 X^-_1) X^-_1 X^-_2 \ve{2}, \qquad L^-X^-_2\ve{2}=
(c_1 +c^2 X^-_2 X^-_1) X^-_1 X^-_2 \ve{2},
\vspace{2mm}\\
\vc{2}L^+=\vc{2} X^+_2 X^+_1(\bar c_2+\bar c^2 X^+_1),\qquad
\vc{2} X^+_2 L^+= \vc{2} X^+_2 X^+_1 (\bar c_1+\bar c^2 X^+_1 X^+_2).
\ea
\]

Substituting this expression into (\ref{MCC"}) after some trivial
evaluations we come to the following relation:
\begin{equation}
\ba{l}
\ds u(u^{-1} u_x)_y =(\mbox{Det}\; u)^{-1}
\pmatrix{ \bar c_2+\bar c^2 (X^+_1)_l, & 0 \cr
        \bar c_1+\bar c^2 (X^+_1 X^+_2)_l , & 0 \cr}
\vspace{3mm}\\
\ds \qquad \qquad \qquad \times \pmatrix{ c_2+c^2 (X^-_1)_r , & c_1 +c^2 (X^-_2 X^-_1)_r \cr
                   0 , &  0 \cr} \mbox{Det}_3.
\ea \label{F}
\end{equation}
In the last expression $\mbox{Det}_3$ satisfy all conditions of (\ref{GJI}),
with the sequence of bases vectors:
\[
\vc{2}, \quad \vc{2} X^+_2, \quad \vc{2} X^+_2 X^+_1.
\]
In this case the summed value of Cartan element $h_1$ is equal to
2, of $h_2$ -- to 0. So with the correct value of numerical factor we
obtain $\mbox{Det}_3=2 \vc{1} K \ve{1}^2 $.

The action of the f\/irst line operator in (\ref{F}) on $(\vc{1} K \ve{1})^2$
leads to the line of the form:
\begin{equation}
2 (\vc{1} K \ve{1})^2 (c_2+2 c^2 \alpha_1 , c_1 +2 c^2 \alpha_{21}) \label{L},
\end{equation}
where following abbreviations are used:
\be
\ba{l}
\ds \bar \alpha_1={\vc{i} X^+_i K \ve{i}\over \vc{i} K \ve{i}},\qquad
\bar \alpha_{12}={\vc{1} X^+_1 X^+_2 K \ve{1}\over \vc{1} K \ve{1}},
\vspace{3mm}\\
\ds \bar \alpha_{21}={\vc{2} X^+_2 X^+_1 K \ve{2}\over \vc{2} K \ve{2}}, \qquad
\alpha_i={\vc{i} K X^-_i \ve{i}\over \vc{i} K \ve{i}},\qquad  i=1,2,
\vspace{3mm}\\
\ds \alpha_{21}={\vc{1} K X^-_2 X^-_1 \ve{1}\over \vc{1} K \ve{1}},\qquad
\alpha_{12}={\vc{2} K X^-_1 X^-_2 \ve{2}\over \vc{2} K \ve{2}}.
\ea\label{NOT}
\ee
Now it is necessary to act with the help of the column operator (\ref{F})
on the line (\ref{L}). The result of this action on scalar factor may be
presented in the form ($\mbox{Det}_2 u=\vc{1} K \ve{1}^2$):
\[
2\pmatrix{ \bar c_2+2 \bar c^2 \bar \alpha_1, & 0 \cr
        \bar c_1+2 \bar c^2 \bar \alpha_{12} , & 0 \cr}
\pmatrix{ c_2+2 c^2 \alpha_1 , & c_1 +2 c^2 \alpha_{21} \cr
                   0 , &  0 \cr}.
\]
The action of the column operator (\ref{F}) on the line (\ref{L}) leads
to additional matrix:
\[
4c^2 \bar c^2 \pmatrix{ (X^+_1)_l \alpha_1 & (X^+_1)_l \alpha_{21} \cr
                (X^+_2 X^+_1)_l \alpha_1 & (X^+_2 X^+_1)_l \alpha_{21} \cr}.
\]
With the help of formulae of Appendix~I the last matrix may be evaluated to
the form:
\[
4 c^2 \bar c^2 (\mbox{Det}\; u)^{-1} u .
\]
Gathering all results together, we obtain f\/inally:
\begin{equation}
u(u^{-1} u_x)_y =2 \pmatrix{ p_1 \bar p_1, & p_2 \bar p_1 \cr
                             p_1 \bar p_2, & p_2 \bar p_2 \cr}+
4 c^2 \bar c^2 (\mbox{Det} \;  u)^{-1} u, \label{B_2}
\end{equation}
where
\[
p_1=c_2+2 c^2 \alpha_1,\qquad \bar p_1=\bar c_2+2 \bar c^2 \bar \alpha_1,\qquad
p_2=c_1+2 c^2 \alpha_{21},\qquad \bar p_2=\bar c_1+2 \bar c^2 \bar \alpha_{12}.
\]

Now we would like to show that the derivatives $(p_{\alpha})_y$ and
$(\bar p_{\alpha})_x$ are functionally dependent on matrix $u$ and
themselves, closing in this way the system of equations of equivalence and
representing it in the form of closed system of equations for $8$ unknown
functions: 4~matrix elements of $u$ and $4$ components of $2$ two-dimensional
spinors $p$, $\bar p$.

Let us follow now the main steps of the necessary calculations.
Using the introduced above technique we have subsequently:
\[
(p_1)_y=2 c^2 (\alpha_1)_y=
{ 2 c^2\over \mbox{Det}\, (u)}\;
\mbox{Det}\pmatrix{ \vc{1} K \ve{1} & \vc{1} K X^-_1\ve{1} \cr
         \vc{1} L^+ K \ve{1} & \vc{1} L^+ K X^-_1 \ve{1} \cr}.
\]
The action of $L^+$ on the state vector $\vc{1}$ is the following:
\[
\vc{1}L^+=\vc{1} X^+_1(\bar c_1-\bar c_2 X^+_2-2 \bar c^2 X^+_2 X^+_1).
\]
Substituting the last expression in the previous equation and
using the f\/irst Jacobi identity for its two f\/irst terms
(linear in $\bar c_1$, $\bar c_2$) we obtain:
\be
\ba{l}
\ds (p_1)_y={ 2 c^2\over \mbox{Det}\,(u)}(\bar c_1 \vc{2} K \ve{2} - \bar c_2 \vc{2} X^+_2
K \ve{2})
\vspace{3mm}\\
\ds \qquad -{ 2 c^2\over \mbox{Det}\, (u)} \; \mbox{Det}
\pmatrix{ \vc{1} K \ve{1} & \vc{1} K X^-_1\ve{1} \cr
\vc{1}X^+_1X^+_2 X^+_1  K \ve{1} & \vc{1}X^+_1X^+_2 X^+_1 K X^-_1 \ve{1} \cr}.
\ea
\label{AE}
\end{equation}
Substituting into the second Jacobi identity (\ref{2JI}) ($p=2$) the f\/irst
one in the form:
\[
\vc{2} K \ve{2}=\mbox{Det}\pmatrix{ \vc{1} K \ve{1} & \vc{1} K X^-_1\ve{1} \cr
         \vc{1} X^+_1 K \ve{1} & \vc{1} X^+_1 K X^-_1 \ve{1} \cr}
\]
we obtain after some trivial transformations equality for two
second order determinants:
\[
\ba{l}
\ds \pmatrix{ \vc{1} X^+_1 K \ve{1} & \vc{1} X^+_1 K X^-_1\ve{1} \cr
\vc{1}X^+_1X^+_2  K \ve{1} & \vc{1}X^+_1X^+_2 K X^-_1 \ve{1} \cr}
\vspace{3mm}\\
\qquad = \pmatrix{ \vc{1} K \ve{1} & \vc{1} K X^-_1\ve{1} \cr
\vc{1}X^+_1X^+_2 X^+_1  K \ve{1} & \vc{1}X^+_1X^+_2 X^+_1 K X^-_1 \ve{1} \cr}.
\ea
\]
Evaluating the last column of the f\/irst determinant with the help of
the f\/irst Jacobi identity:
\[
\ba{l}
\ds \vc{1} X^+_1 K X^-_1\ve{1}={\vc{2} K \ve{2}+\vc{1} X^+_1 K \ve{1} \vc{1} K
X^-_1\ve{1}\over \vc{1} K \ve{1}},
\vspace{3mm}\\
\ds \vc{1} X^+_1 X^+_2 K X^-_1\ve{1}={\vc{2} X^+_2 K \ve{2}+\vc{1} X^+_1 X^+_2 K
\ve{1} \vc{1} K X^-_1\ve{1}\over \vc{1}  K \ve{1}}
\ea
\]
we obtain for it:
\[
\bar \alpha_1 \vc{2} X^+_2 K \ve{2}-\bar \alpha_{12} \vc{2} K \ve{2}.
\]
Finally we have:
\begin{equation}
(p_1)_y={ 2 c^2\over \mbox{Det}\,(u)}(u_{11}\bar p_2 -u_{21}\bar p_1),\qquad
(p_2)_y={ 2 c^2\over \mbox{Det}\,(u)}(u_{12 }\bar p_2 -u_{22}\bar p_1). \label{UB_2}).
\end{equation}
So (\ref{B_2}), (\ref{UB_2}) and the same system for derivatives of $(\bar p)_x$
is the closed system of identities or $B_2(1,0;2,2;c^2,\bar c^2)$
exactly integrable system connected with the $B_2$ semisimple serie.
To the best of our knowledge this system was not mentioned in literature
before.

>From the physical point of view the exactly integrable system (\ref{B_2}),
(\ref{UB_2}) may be considered as a model of interacting charge ${1\over 2}$
particle $(\bar p, p)$ with scalar-vector neutral f\/ield $u$.

Putting $ c^2=\bar c^2=0$, we come back to nonabelian Toda lattice system for
single matrix valued unknown function $u$.

\subsubsection{(0,1) grading}

Generators $L^{\pm}$ contain only the components with $\pm 1$ graded
indexes and have the form:
\[
\ba{l}
\ds L^+=\bar d_1 X^+_2+\bar d_2 [X^+_1,X^+_2]+{1\over2} \bar d_3 [X^+_1[X^+_1,
X^+_2]],
\vspace{3mm}\\
\ds L^-=d_1 X^-_2+d_2 [X^-_2,X^-_1]+{1\over 2} d_3 [ X^-_1[X^-_1,X^-_2]].
\ea
\]
With respect to transformation of $1$ -- red group $A_1$ functions
$d_i(x)$, $(\bar d_i(y))$ are components of three dimensional $A_1$ vectors.

The object of investigation is two dimensional matrix $u$ in the basis of
the second fundamental representation of $B_2$ algebra. The main equation
(\ref{MCC"}) conserves its form. The action of $L^{\pm}$ operators on the
basis vectors have now the form\footnote{Four basis vectors of the f\/irst fundamental of
the $C_2$ algebra are the following: $ \ve{1}$, $X^-_1 \ve{1}$,
$X^-_2 X^-_1 \ve{1}$, $X^-_1 X^-_2 X^-_1 \ve{1}$.}:
\[
\ba{l}
L^-\ve{1}=(d_2-d_3 X^-_1) , \qquad L^-X^-_1\ve{1}=
(d_1 -d_2 X^-_1) X^-_2 X^-_1 \ve{1},
\vspace{2mm}\\
\vc{2}L^+=\vc{2} X^+_2 X^+_1(\bar d_2-\bar d_3 X^+_1),\qquad
\vc{2} X^+_2 L^+= \vc{2} X^+_2 X^+_1 (\bar d_1-\bar d_2 X^+_1 X^+_2).
\ea
\]

Substituting this expression into (\ref{MCC"}), keeping in mind that $\mbox{Det}_3$
satisfy all conditions~of (\ref{GJI}), after some trivial
evaluations we come to the following relation $(\mbox{Det}_3=\vc{1} K \ve{1})$:
\begin{equation}
\ba{l}
\ds u(u^{-1} u_x)_y =(\mbox{Det}\, u)^{-1}
\pmatrix{ \bar d_2-\bar d_3 (X^+_1)_l, & 0 \cr
        \bar d_1-\bar d_2 (X^+_1)_l , & 0 \cr}
\vspace{3mm}\\
\ds \qquad \qquad \times \pmatrix{ d_2-d_3 (X^-_1)_r , & d_1 -d_2 ( X^-_1)_r \cr
                   0 , &  0 \cr} \vc{1} K \ve{1}.
\ea \label{FF}
\end{equation}
Not cumbersome transformation leads the last expression to the f\/inally form:
\begin{equation}
(u^{-1} u_x)_y =(\mbox{Det}\,  u)^{-1} u^{-1}
\pmatrix{ \bar d_2, & -\bar d_3  \cr
                   \bar d_1, & -\bar d_2 \cr} u
\pmatrix{ d_2, & d _1  \cr
          -d_3, & -d_2 \cr}.  \label{B21}
\end{equation}
(\ref{B21}) is nonabelian Toda chain for $B_2$ algebra with $(0,1)$
grading. To the best of our knowledge it was not considered before.

System (\ref{B21}) is form-invariant with respect to transformation
$u\to \bar g(y) u g(x)$, with the help of which it is possible to
evaluate matrices depending on $x$, $y$ arguments to constant values. We omit
here the question about the possible canonical forms of the system
$B_2(0,1;1,1;\bar d,d)$ (\ref{B21}).

\subsection{The case of {\mathversion{bold}$G_2$} algebra}

As it is possible to expect, this case is the most cumbersome. It corresponds to
the choise $p=3$ in (\ref{RS}). Firstly, we will consider the case of
$(0,1)$ grading as the most simple one. It is connected with the 7-th
dimensional f\/irst fundamental representation of $G_2$ algebra (group).
The second one connected with $(1,0)$ grading is $14$-th dimensional.

\subsubsection{(0,1) grading}

In this case $L^{\pm}$ may contain the components $\pm 1$, $\pm 2$
graded subspaces and have the form:
\[
\ba{l}
\ds L^+= \bar d_1 X^+_2+ \bar d_2 [X^+_1,X^+_2]+{1\over 2}\bar d_3 [X^+_1[X^+_1,
X^+_2]]
\vspace{3mm}\\
\ds \qquad +{1\over 6}\bar d_4 [X^+_1[X^+_1[X^+_1,X^+_2]]]+{1\over 3} \bar d^2
[X^+_2[X^+_1[X^+_1[X^+_1,X^+_2]]]],
\ea
\]
$L^-=(L^+)^T$, where $T$ sign of transposition; with
simultaneously exchanging all coef\/f\/icients $\bar d\to d$. This
operation we will call as ``hermitian conjugation".

Four coef\/f\/icient functions $d_i$, $\bar d_i$ on the generators of the $\pm 1$
graded subspaces in $L^{\pm}$ are united to the ${3\over 2}$ multiplate,
with respect to gauge transformation initiated by group elements $g_0(x)$,
$\bar g_0(y)$ belonging to the f\/irst red group.

The f\/irst fundamental representation of $G_2$ algebra is $7$-th dimensional
with the basis vectors:
\[
\ba{l}
\ve{1}, \quad X^-_1\ve{1}, \quad X^-_2X^-_1\ve{1}, \quad  X^-_1X^-_2X^-_1\ve{1}, \quad X^-_1X^-_1
X^-_2X^-_1\ve{1},
\vspace{2mm}\\
X^-_2X^-_1X^-_1X^-_2X^-_1\ve{1}, \quad X^-_1X^-_2X^-_1X^-_1X^+_2X^-_1\ve{1}.
\ea
\]

The action of the operators $L^{\pm}$ on $A_1$ basis of $u$ matrix is as
follows:
\[
\ba{l}
\ds \vc{1} L^+=\vc{1} X^+_1X^+_2(\bar d_2 -\bar d_3 X^+_1+{1\over 2} \bar d_4
X^+_1X^+_1- \bar d^2 X^+_1X^+_1X^+_2),
\vspace{3mm}\\
\ds \vc{1} X^+_1L^+=\vc{1} X^+_1X^+_2(\bar d_1 - \bar d_2 X^+_1+{1\over 2}
\bar d_3 X^+_1X^+_1-\bar d^2 X^+_1X^+_1X^+_2X^+_1).
\ea
\]
The action of the operator $L^-$ on $A_1$ basis from the left may
be obtained from the last formulae with the help of ``hermitian
conjugation'':
\[
L^-\ve{1}= (\vc{1} L^+)^T,\qquad  L^-X^-_1\ve{1}= (\vc{1} X^+_1L^+)^T,\qquad
\bar d\to d.
\]

As in the previous sections the result of calculation of the main determinant
(\ref{MCC"}) it is possible to present in the form of the product of column
operator on the line one applied to the highest vector $\vc{1} K \ve{1}^2$
of the $(2,0)$ representation of $G_2$ algebra (see Appendix~II).
The line operator form is the following
\[
\ba{l}
\ds \Bigr [d_2-d_3X^-_1+{1\over 2}d_4(X^-_1)^2-{1\over 4} d^2(2X^-_1X^-_2-3X^-_2X^-_1)
X^-_1,
\vspace{3mm}\\
\ds \qquad d_1-d_2X^-_1+{1\over 2}d_3(X^-_1)^2-{1\over 4}d^2X^-_1(2X^-_1X^-_2-3X^-_2
X^-_1)X^-_1\Bigr],
\ea
\]
where $X^-_i\equiv (X^-_i)_r$. Nonusual (compared with the previous examples)
form of the coef\/f\/icient on $d^2$ term is the prise for $p=3$ in (\ref{RS})
in the case of $G_2$ algebra.
The column operator is obtained from the line one with the
help introduced above rules of the ``hermitian conjugation''.

For rediscovering of last symbolical expression up to the form of usual
$2\times 2$ matrix let us introduce two dimensional column vector $\bar q$
the result of the action of the column operator on the highest vector
$\vc{1} K \ve{1}^2$ divided by itself. The same in the line case will be
denoted as $q$. Explicit expressions for the line
components of $q$ have the form:
\be\label{Q}
\ba{l}
\ds q_1=\left(d_2+{1\over 3}d^2\alpha_{112}\right)-
2\left(d_3+{2\over 3}d^2\alpha_{12}\right)\alpha_1+\left(d_4+2\alpha_2d^2\right)\alpha_1^2,
\vspace{3mm}\\
\ds q_2=\left(d_1+{1\over 3}d^2\alpha_{1112}\right)-
2\left(d_2+{1\over 3}d^2\alpha_{112}\right)\alpha_1+
\left(d_3+{2\over 3}d^2\alpha_{12}\right)\alpha_1^2
\ea
\ee
and with the help of ``hermitian conjugation'' corresponding expressions for
the components for the column $\bar q$.

The result of the action of line operator on the highest vector in connection
with all said above is equal to the numerical line vector $\vc{1} K \ve{1}^2
(q_1, q_2)$. The action of the column operator on it may be devided on two
steps: the action on the scalar factor $\vc{1} K \ve{1}^2 $, with the f\/inally
matrix $\vc{1} K \ve{1}^2 \bar q q$ (multiplication by the law the column on
the line) and the terms with partial mutual dif\/ferentiation of the scalar and
the lines factors. All formulae for concrete calculation of such kind the
reader can f\/ind in Appendix~II. It is necessary to pay attention to the fact,
that $X^+_2 q_i=X^-_2 \bar q_i=0$, which one can check without any dif\/f\/iculties
with the help of formulae of the Appendix~I.

Gathering all these results we obtain the equation of equivalence for $u$
function:
\begin{equation}
u(u^{-1}u_x)_y={\det}^{-1}(u)\sum_{i,j,k,l} u_{ij} u_{kl} \epsilon_{ik}
\epsilon_{jl}\bar p^{ik} p^{jl}+4 d^2\bar d^2(\mbox{Det}\, (u))^{-1} u,\label{BE}
\end{equation}
where $u_{ij}$ elements of the matrix $u$, $\epsilon_{ij}$ symmetrical
tensor of the second rank with the components $\epsilon_{12}=\epsilon_{21}=-1$,
$\epsilon_{11}=\epsilon_{22}=1$, $\bar p^{ij}$, $p^{ij}$ are two-dimensional
column and line vectors correspondingly with the components (the law
of multiplication is the column on the line):
\[
\ba{l}
\ds p^{11}=\left(d_2+{1\over 3} d^2\alpha_{112},d_1+{1\over 3} d^2\alpha_{1112}\right),
\qquad p^{22}=\left(d_4+2d^2\alpha_2,d_3+{2\over 3} d^2\alpha_{12}\right),
\vspace{3mm}\\
\ds p^{12}=p^{21}=\left(d_2+{1\over 3} d^2\alpha_{112},d_3+{2\over 3} d^2\alpha_{12}\right).
\ea
\]

It remains only to f\/ind the derivatives $(\bar p_{ij})_x$, $(p_{kl})_y$ and convince ourselves
that together with the (\ref{BE}) they compose the closed system of equations of equivalence
or exactly integrable $G_2(0,1;2,2;\bar d^2,d^2)$ system.

Four components of  $p^{22}$, $p^{11}$ with respect to transformation of the
f\/irst red algebra compose the ${3\over 2}$ spin-multiplet. So it will be
suitable to redenote them by single four-dimensional symbol $p_i$. And the
same for ``hermitian conjugating'' values $\bar p_i$.

Let us follow the calculation of $(\bar p_4)_x=2\bar d^2 (\bar \alpha_2)_x$.
The calculation of this the derivative do not dif\/ferent from the
corresponding computations of Section~3 (see (\ref{AR}) and (\ref{ARR})).
We have consequently:
\begin{equation}
(\bar \alpha_2)_x=\vc{2} K \ve{2})^{-2} \pmatrix{
\vc{2} K \ve{2}, & \vc{2} K L^- \ve{2} \cr
        \vc{2} X^+_2 K \ve{2}, &  \vc{2} X^+_2 K L^- \ve{2} \cr}.
\label{ARm}
\end{equation}
With the help of the technique used many times before we evaluate the last
expression to:
\[
\ba{l}
(\bar \alpha_2)_x=L^-_r(X^+_2)_l \ln \vc{2} K \ve{2}
 =\left[d_1-d_2 X^-_1+{1\over 3} d_3 (X^-_1)^2\right.
\vspace{3mm}\\
\ds \qquad \left. -{1\over 6} d_4 (X^-_1)^3+
d^2([[[X^-_2,X^-_1]X^-_1]X^-_1]-X^-_2(X^-_1)^3)\right] \theta_2.
\ea
\]
Using with respect to the last expression formulae of Appendix~I we come to
the system of equations of equivalence for $\bar p$ components of
${3\over 2}$ multiplet:
\be \label{BLE}
\ba{l}
\ds (\bar p_4)_x={2\bar d^2\over \mbox{Det}^2 (u)}\left(p_1 u_{11}^3-3p_2 u^2_{11} u_{12}+
3p_3 u_{11} u^2_{12}-p_4 u^3_{12}\right),
\vspace{3mm}\\
\ds (\bar p_3)_x={2\bar d^2\over \mbox{Det}^2 (u)}\left(p_1 u_{11}^2 u_{21}-p_2 \left(u^2_{11}
u_{22}+ 2u_{11} u_{21} u_{12}\right)\right.
\vspace{3mm}\\
\ds\phantom{(\bar p_2)_x=} \left. +p_3 \left(2u_{11} u_{12}u_{21}+u^2 _{12}u_{21}\right)-p_4 u^2_{12}u_{22}\right),
\vspace{3mm}\\
\ds (\bar p_2)_x={2\bar d^2\over \mbox{Det}^2 (u)}
\left(p_1 u_{11} u^2_{21}-p_2 \left(u^2_{21} u_{12}+ 2u_{11} u_{21} u_{22}\right)\right.
\vspace{3mm}\\
\ds \phantom{(\bar p_2)_x=} \left. +p_3 \left(2u_{22} u_{12}u_{21}+u^2 _{22}u_{11}\right)-p_4
u^2_{22}u_{12}\right),
\vspace{3mm}\\
\ds (\bar p_1)_x={2\bar d^2\over \mbox{Det}^2 (u)}\left(p_1 u_{21}^3-3p_2 u^2_{21} u_{22}+
3p_3 u_{21} u^2_{22}-p_4 u^3_{22}\right).
\ea
\ee

And corresponding system for derivatives $p_y$, which can be obtained from
(\ref{BLE}) with the help of ``hermitian conjugation''.

The symmetry of the constructed exactly integrable $G_2(0,1;2,2;\bar d^2,d^2)$
system (\ref{BE}), (\ref{BLE}) is higher than any possible espectations.

>From the physical point of view this system may be considered as the
interuction of charge ${3\over 2}$ spin particle ($p,\bar p$) with
neutral scalar-vector f\/ield $u$.

\subsubsection{(1,0) grading}

In this case  $L^{\pm}$ may contain the components $\pm 1$, $\pm 2$, $\pm 3$
graded subspaces and have the form:
\[
\ba{l}
L^+=\bar c_1 X^+_1+ \bar c_2 [X^+_1,X^+_2]+\bar c^2 [X^+_1[X^+_1,X^+_2]]
\vspace{2mm}\\
\ds \qquad + \bar c^3_1 [X^+_1[X^+_1[X^+_1,X^+_2]]]+\bar c^3_2 [X^+_2[X^+_1[X^+_1[X^+_1,
X^+_2]]]],
\ea
\]
$L^-=(L^+)^T$, where $T$ sign of transposition
($(X^+_i)^T=X^-_i)$; with simultaneously exchange of all
coef\/f\/icients $\bar c\to c$. This operation was called as
``hermitian conjugation'' in the previous subsection and we
conserve here this notation.

As always we begin from the equation of equivalence for two dimensional
matrix $u$ connected with the second simple root of $G_2$ algebra.
For the decoded of universal equation~(\ref{MCC"}) it is necessary
the knowledge of the action of $L^{\pm}$ on the basis. We represent below only
part of basis vectors of the second fundamental ($14$-th dimensional)
representation of $G_2$ algebra:
\[
\ba{l}
\ve{2}, \ X^-_2\ve{2}, \ X^-_1X^-_2\ve{2}, \  X^-_1X^1_2X^-_2\ve{2}, \
X^-_1X^-_1X^-_1X^-_2\ve{2}, \ X^-_2X^-_1X^-_1X^-_2\ve{2},
\vspace{2mm}\\
X^-_2X^-_1X^-_1X^-_1X^-_2\ve{2}, \ X^-_1X^-_2X^-_1X^-_1X^-_2\ve{2}, \
X^-_1X^-_2X^-_1X^-_1X^-_1X^-_2\ve{2}.
\ea
\]
The main equations (\ref{I}) are obviously invariant with the respect to
the gauge transformation iniciated  by $g_0(x)$, $\bar g_0(y)$ elements of the
red algebra of the second simple root. With respect to this transformations
two coef\/f\/icients of zero $(c^1,\bar c^1)$ and third $(c^3,\bar c^3)$
order graded subspaces are transformed as spinor (anti-) multiplets; $c^2$,
$\bar  c^2$ are the scalar ones. With the help of such transformation it is always
possible to satisfy the condition $c^3_2=\bar c^3_2=0$ (what is essential
simplif\/ied the calculation) and reconstruct the general case at the f\/inal
step using invariance condition.

The action of the $L^{\pm}$ operators on the basis states of the second red
algebra has the form:
\[
\ba{l}
\vc{2} L^+=\vc{2} X^+_2X^+_1(-\bar c^1_2 +\bar c^2 X^+_1-\bar c^3_1 X^+_1X^+_1
+\bar c^3_2 (2X^+_1X^+_1X^+_2-3X^+_1X^+_2X^+_1),
\vspace{2mm}\\
\vc{2} X^+_2L^+=\vc{2} X^+_2X^+_1(\bar c^1_1 +\bar c^2 X^+_1X^+_2+
(X^+_1X^+_1X^+_2-3X^+_1X^+_2X^+_1)(\bar c^3_1-\bar c^3_2 X^+_2).
\ea
\]
The action of the operator $L^-$ on $A_1$ basis from the left may be obtained
from the last formulae with the help of ``hermitian conjugation'':
\[
L^-\ve{2}= (\vc{2} L^+)^T,\qquad  L^-X^-_1\ve{2}= (\vc{2} X^+_1L^+)^T,\qquad
\bar c\to c.
\]

Taking into account arguments of the Appendix~II the result of the
calculation of determinant of the third order (\ref{MCC"}) may be
presented in the operator column on line form, acting on the
highest vector of $(4,0)$ representation ($3 \vc{1} K \ve{1}^4$)
of $G_2$ algebra.

The line (``hermitian conjugating'' column) operators has the form (in this
expression we put $c^3_2=\bar c^3_2=0$):
\[
\left(-c^1_2 +c^2 X^-_1-{1\over 2}\bar c^3_1 (X^-_1)^2,\quad
c^1_1 +c^2 X^-_2X^-_1+{1\over 8}c^3_1
(X^-_2X^-_1X^-_1-6X^-_1X^-_2X^-_1)\right).
\]
Further calculations are on the level of accurate application of
dif\/ferentiation rules and combination terms of the same nature.
Equation of equivalence for $u$ function have the f\/inal form:
\begin{equation}
u(u^{-1}u_x)y=3 \, {\det}^{{1\over 3}}(u) \bar p^1 p^1+12 \, {\det}^{-{1\over 3}}(u)
\bar p^2 p^2 u+18 \,{\det}^{-1}(u) (u\bar c^3) (c^3 u),
\label{HV}
\end{equation}
where $p^1$ is the spinor with the components $p^1=(-c^1_2+4c^2\alpha_1
-6c^3_1\alpha_1^2, c^1_1+4c^2
\alpha_{21}-c^3_1(\alpha_{121}+2\alpha_1\alpha_{21}))$; scalar $p^2=
c^2-3c^3_1\alpha_1$ and corresponding expressions for bar values.

We present the system of equivalence equations without any further comments:
\begin{equation}
(\bar p^2)_x=-3 \, {\det}^{-{2\over 3}} \sum_{i,j,k,l}\bar c^3_i u_{ij}\epsilon_{kl}
p^1_l,\qquad (\bar p^1_i)_x={\det}^{-{2\over 3}}\bar p^2 \sum_{j,k,l} u_{ij}
\epsilon_{k,l}p^1_l,\label{LLL}
\end{equation}
where $\epsilon_{k,l}=-\epsilon_{l,k}$ antisymmetrical tensor of the second
rank $\epsilon_{1,2}=-\epsilon_{2,1}=1$.
And, of, course the corresponding system with the derivatives $p^1_y$, $p^2_y$.

Physical interpretation of the last system may be connected with spinor
particle interacting with charged scalar $(p^2,\bar p^2)$ and neutral
scalar-vector f\/ields in two dimensions.

\section{Concluding remarks}

In some sense in the present paper the initial idea of Sofus Lie
to introduce continuous groups as powerful apparatus for solving
the dif\/ferential equations is realized.

On the examples of semisimple groups of second order we have decoded this
idea and described  explicitly exactly integrable systems whose general
solutions can be obtained with the help and in the terms of
group representation theory. We have no doubts (and partially can prove
this) that the same construction is applicable to the case of arbitrary
Lie groups and hope to prove this statement completely or to see the proof in
the literature in the nearest future.

\subsection*{Acknowledgements}

Author is indebted to the Instituto de Investigaciones en Matem'aticas
Aplicadas y en Sistemas, UNAM for beautiful conditions for his work.
Author freundly thanks N.~Ata\-ki\-shiyev for permanent discussions in the
process of working on this paper and big practical help.

This work was done under partial support of Russian Foundation of Fundamental
Researches (RFFI) GRANT.

\renewcommand{\theequation}{{\rm I}.\arabic{equation}}
\setcounter{equation}{0}

\section*{Appendix~I}

The formulae below are the general ones and have in their foundation the
f\/irst Jacobi identity only.

Let us def\/ine:
\[
\theta_j=\prod_{i=1}^r (\vc{i} G \ve{i})^{-K_{ji}}.
\]
As a result of dif\/ferentiation of $\ln \theta_i$, we obtain:
\begin{equation}
(X^-_q)_r \theta_i=-\theta_i  K_{iq} \alpha_q,\qquad (X^+_q)_l \theta_i=
-\theta_i K_{iq} \bar \alpha_q, \label{AI1}
\end{equation}
\begin{equation}
(X^-_q)_r \bar \alpha_i=\delta_{q,i} \theta_i,\qquad (X^+_q)_l \alpha_i=
\delta_{q,i} \theta_i. \label{AI2}
\end{equation}

In the case of the second order algebras:
\begin{equation}
\theta_1={\vc{2} G \ve{2}\over \vc{1} G \ve{1}^2},
\qquad \theta_2={\vc{1} G \ve{1}^p\over \vc{1} G \ve{1}^2}.\label{AI3}
\end{equation}

\renewcommand{\theequation}{{\rm II}.\arabic{equation}}
\setcounter{equation}{0}

\section*{Appendix II}

Let us consider the determinant of the third order the matrix entiries
of which are coinsided with the matrix elements of $G_2$ group element $K$
taken between the bra and the ket three dimensional bases:
\begin{equation}
\vc{1},\ \vc{1} X^+_1,\ \vc{1} X^+_2X^+_1X^+_1X^+_2X^+_1,\
\ve{1}, \ X^-_1\ve{1}, \ X^-_2X^-_1X^-_1X^-_2X^-_1\ve{1}.\label{AII}
\end{equation}

Acting on such determinant by generator $(X^+_2)_r$ and taking
$(X^-_1X^-_1)_r$ out of its sign we come to the following ket basis:
\[
\ve{1}, X^-_1\ve{1}, X^-_2X^-_1\ve{1}
\]
which in connection with the (\ref{GJI}) tell us that the initial $\mbox{Det}_3$
(up to the terms anihilated by generators of the positive simple roots from
right and negative ones from the left) belongs
to $(2,0)$ ($Vh_1=2$, $Vh_2=0$) representation of $G_2$ group. For
initial determinant $Vh_1=1$, $Vh_2=0$. Each basis vector (see (\ref{high})) may be
obtained with
consequent application of the lowering operators to the higest vector
($\vc{1} K \ve{1}^2$ in the present case). There are two possibility to
combination of the lowering operators:
\[
( (AX^-_2X^-_1+BX^-_1X^-_2)X^-_1 )_r
\]
and the same expression from the left combination of the raising
generators. The condition that $\mbox{Det}_3$ is anihilated by
generators $(X^+_1)_r(X^-_1)_l$, which is a direct corollary of
the structure of the bra and ket basises, allow to f\/ind relation
between the constants $3A+2B=0$ and obtain the expression used in
the main text (\ref{Q}) and above. We obtain the following value
for $\mbox{Det}_3$ in basis (\ref{AII}):
\[
\ba{l}
\ds \mbox{Det}_3={1\over 16}((2X^-_1X^-_2-3X^-_2X^-_1)X^-_1)
\vspace{3mm}\\
\ds \phantom{\mbox{Det}_3=} \times (X^+_1(2X^+_2X^+_1- 3X^+_1X^+_2))
\vc{1} K \ve{1}^2+\vc{1} K \ve{1}.
\ea
\]

Below we present necessary formulae for calculation of (\ref{BE}).
We restrict ourselves by (11) component of it.
All ``mixed'' terms may be gothered in the following form:
\be
\ba{l}
\ds -{\bar d^2\over 4}\left[2(X^+_1X^+_2X^+_1 q_1)-3(X^+_2X^+_1X^+_1 q_1)-8\bar
\alpha_1(X^+_2X^+_1 q_1)-8\bar \alpha_1 (X^+_1 q_1)\right]
\vspace{3mm}\\
\ds \qquad - \bar d_3 (X^+_1 q_1)+2\bar d_4
\bar \alpha_1(X^+_1 q_1)+{\bar d_4\over 2} ((X^+_1)^2 q_1).
\ea
\label{AII1}
\end{equation}
Using the def\/inition of vector $q$ (\ref{Q}) and formulae of Appendix~I,
we obtain:
\[
\ba{l}
(X^+_1 q_1)=2\theta_1(p^{22}_1\alpha_1-p^{22}_2)\equiv 2\theta_1 P,\quad
X^+_2 P=0,\quad (X^+_2X^+_1 q_1)=2\theta_1 \bar \alpha_2 P,
\vspace{2mm}\\
(X^+_1X^+_1 q_1)=2\theta_1^2 p^{22}_1-4\theta_1 \bar \alpha_1 P,\quad
(X^+_1X^+_2X^+_1 q_1)=2\theta_1(\bar \alpha_{21}-2 \bar \alpha_1 \bar \alpha_2
)P+2\theta_1^2 \bar \alpha_2 p^{22}_1,
\vspace{2mm}\\
(X^+_2X^+_1X^+_1 q_1)=4\theta_1^2 \bar \alpha_2 p^{22}_1+4d^2 \theta_1^2
\theta_2-4\theta_1 \bar \alpha_1 \bar \alpha_2 P-4\theta_1 \bar \alpha_{12} P.
\ea
\]

\label{leznov-lp}


\begin{thebibliography}{99}
\footnotesize

\bibitem{leznov:LG} Leznov A.N., {\it TMF}, 1980, V.42, 343--355.

\bibitem{leznov:ls0} Leznov  A.N. and Saveliev M.V., Group-Theoretical
Methods for Integration of Nonlinear Dynamical Systems, Vol.~15
of Progress in Physics, Birkhauser-Verlag, Basel, 1992.

\bibitem{leznov:LM}
Leznov A.N., Graded Lie Algebras, Representation Theory,
Integrable Mappings and Systems, UNAM Preprint 83, 1998.\\
Leznov A.N., The Exactly Integrable Systems Connected with
Arbitrary Semisimple Algebras. The Abelian case, UNAM 85, 1998.


\bibitem{leznov:l} Leznov A.N., Two-Dimensional Ultra-Toda Integrable
Mappings and Chains (Abelian Case), hep-th/9703025.

\bibitem{leznov:GC} Gelfand I. and Retakh V., {\it Funct. An. Appl.}, 1991, V.25, N~2, 91--102.\\
Gelfand I. and Retakh V., {\it Funct. An. Appl.}, 1992, V.26, N~4, 1--20.\\
Etingof P., Gelfand I. and Retakh V., {\it Math. Research Letters}, 1997, V.4, 413--425.

\bibitem{leznov:l1} Leznov A.N., Proceedings of International Seminar ``Group
Methods in Physics'', Zvenigorod, 24--26 November 1982, ed. M.A.~Markov,
Gordon-Breach, New-York, 1983, 443--457.

\end{thebibliography}
\end{document}